\documentclass[12pt]{article}

\usepackage{amssymb}
\usepackage[dvips]{graphicx}
\usepackage[dvips]{psfrag}

\newcommand {\beq}{\begin{equation}}
\newcommand {\eeq}{\end{equation}}
\newcommand {\beqa}{\begin{eqnarray}}
\newcommand {\eeqa}{\end{eqnarray}}
\newcommand {\n}{\nonumber \\}

\newcommand {\del}{\partial}

\def\pa{\partial}

\begin{document}
\setlength{\oddsidemargin}{0cm}
\setlength{\baselineskip}{7mm}

\begin{titlepage}
\renewcommand{\thefootnote}{\fnsymbol{footnote}}
\begin{normalsize}
\begin{flushright}
\begin{tabular}{l}
OU-HET 457 \\
October 2003
\end{tabular}
\end{flushright}
  \end{normalsize}

~~\\

\vspace*{0cm}
    \begin{Large}
       \begin{center}
         {A Note on Hamilton-Jacobi Formalism \\and D-brane Effective Actions}
       \end{center}
    \end{Large}
\vspace{1cm}

\begin{center}
           Matsuo S{\sc ato}\footnote
            {
e-mail address : 
machan@het.phys.sci.osaka-u.ac.jp}
           {\sc and}
           Asato T{\sc suchiya}\footnote
           {
e-mail address : tsuchiya@het.phys.sci.osaka-u.ac.jp }\\
      \vspace{1cm}
       
        {\it Department of Physics, Graduate School of  
                     Science}\\
               {\it Osaka University, Toyonaka, Osaka 560-0043, Japan}\\
      
\end{center}

\hspace{5cm}

\begin{abstract}
\noindent
We first review the canonical formalism with general space-like hypersurfaces
developed by Dirac by rederiving the Hamilton-Jacobi equations
which are satisfied by on-shell actions defined on such hypersurfaces.
We compare the case of gravitational systems with that of the flat space.
Next, we remark as a supplement to our previous results that
the effective actions of D-brane and M-brane given by arbitrary
embedding functions are on-shell actions of supergravities.
\end{abstract}
\vfill
\end{titlepage}
\vfil\eject

\setcounter{footnote}{0}

\section{Introduction}
\setcounter{equation}{0}
\renewcommand{\thefootnote}{\arabic{footnote}}
We showed in \cite{ST, ST2} that the effective actions of D-brane and M-brane 
are on-shell actions of supergravities. We derived the Hamilton-Jacobi (H-J)
equations of supergravities, which are satisfied by on-shell actions,
regarding a radial direction as time, and solved
those equations. We also found that these solutions to the H-J equations
are the on-shell 
actions around the supergravity solutions that are conjectured to be dual to 
various gauge theories, which in particular include
noncommutative super Yang Mills. In the gauge/gravity correspondence, 
the on-shell actions 
in gravities are considered to be generating functionals of 
correlation functions in gauge theories. Therefore, our results 
in \cite{ST, ST2} should be useful for going beyond the AdS/CFT correspondence
and studying the more general gauge/gravity 
correspondence.   

However, it is not clear why the D-brane effective actions which includes
the all-order contributions in the $\alpha'$ expansion
are obtained within
supergravities, which are just the lowest order approximation for string 
theories in the $\alpha'$
expansion. In order to clarify this reason, we must
establish the exact correspondence between
our calculations and the derivations of the D-brane effective action
in string theory.
On one hand, the worldvolumes of the D-brane effective actions
in our solutions are fixed-time hypersurfaces, since the ordinary 
H-J formalism gives on-shell actions defined on the boundary
hypersurfaces specified by the final time. On the other hand, the worldvolumes
of D-branes in string theories are defined by embedding functions which can
specify arbitrary hypersurfaces in the target space \cite{D-brane}. 
Therefore, before
trying to establish the above correspondence, 
we should first investigate whether 
the D-brane effective action whose worldvolume is defined by such
embedding functions is an on-shell action of supergravity or not.

Hence, this brief note 
is concerned with the on-shell action that is obtained
by substituting into the action the classical solution which satisfies
a boundary condition given on not a fixed-time hypersurface but 
an general hypersurface. (See Fig.1.)
We need a generalization of 
the H-J formalism that gives such on-shell actions.
Indeed, this generalization was studied by Dirac around 
fifty years ago \cite{Dirac}. 
The generalization in gravitational 
systems is much simpler. In fact, by performing a general coordinate
transformation that transforms the general hypersurface defined by the
embedding functions to a fixed-time hypersurface, one can show that
the answer to the above question 
is positive. However,
it is interesting and useful for further developments 
to compare the case of gravitational system 
with that of the
flat space, so that we will give a heuristic argument below. 
We will first derive 
Dirac's result in the flat space in a different way, and next
make the comparison. Finally, we remark that 
the effective actions of D-brane and M-brane given by arbitrary
embedding functions are on-shell actions of supergravities

\begin{figure}[htbp]
\begin{center}
\psfrag{aa}{$x^1 \!\! ,\, x^2 \!\! ,\, x^3$}
\psfrag{ab}{$x^0$}
\psfrag{ba}{$x^1 \!\! ,\, x^2 \!\! ,\, x^3$}
\psfrag{bb}{$x^0$}
\psfrag{ca}{$X^M(\sigma^i)$}
\includegraphics[height=5cm, keepaspectratio, clip]{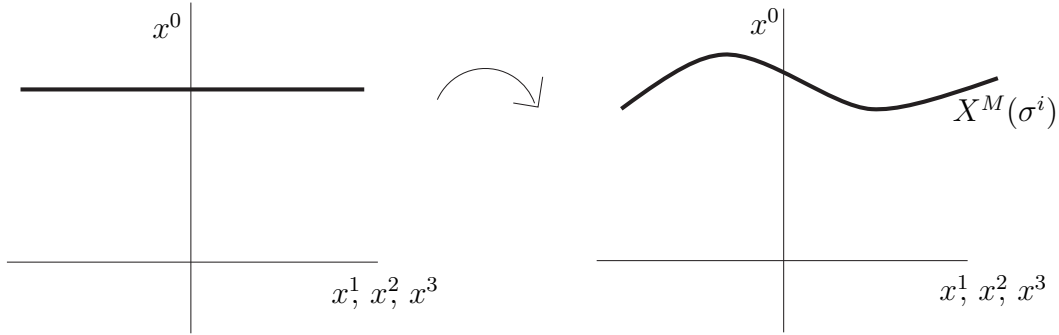}
\end{center}
\caption{Generalization of a boundary surface}
\label{boundary}
\end{figure}


\section{Introduction of dynamical coordinates }
\setcounter{equation}{0}
As we mentioned, we pay attention to 
on-shell actions with the boundary values of fields given on 
general space-like hypersurfaces. Let us consider a four-dimensional scalar 
field theory in the flat space as an example. The action is given by   
\beq
I = -\int d^4x \left(\frac{1}{2}\eta^{MN} \del_M \phi(x) \del_N \phi(x) + V(\phi(x)) \right),  \label{scalar action 0}
\label{I}
\eeq
where $\eta_{MN}=\mbox{diag}(-1,1,1,1)$. The equation of motion is
\beq
\eta^{MN} \del_M \del_N \phi = V'(\phi). \label{e.o.m. of scalar}
\eeq
The space-like hypersurface which specifies the boundary is parametrized by 
the $X^M(\sigma^i)$, where $i=1,2,3.$ (See Fig.1.)
In principle, we can obtain the solution 
$\bar{\phi}(x)$ to the equation of motion (\ref{e.o.m. of scalar})
which satisfies 
a boundary condition 
\beq
\bar{\phi}(X^M(\sigma^i))=\Phi(X^M(\sigma^i)). \label{boundary condition}
\eeq
By substituting this solution into $I$, we obtain the on-shell action in which 
we are interested,
\beq
S(X, \Phi(X)) = -\int d^4x \left(\frac{1}{2}\eta^{MN} \del_M \bar{\phi}(x) \del_N \bar{\phi}(x) + V(\bar{\phi}(x)) \right). \label{scalar on-shell action }
\eeq
It is in general difficult to solve the equation of motion in the above situation and to obtain the on-shell action directly. Instead, we seek for the Hamilton-Jacobi equations, which are the differential equations satisfied 
by the on-shell action of this kind. 

In order to develop the generalized
Hamilton-Jacobi formalism, we consider an action,
\beq
\tilde{I} = -\int d^4 \sigma \mbox{det} \left( \frac{\del x(\sigma)}{\del \sigma} \right) \left( \frac{1}{2}\eta^{MN} \frac{\del \sigma^{\alpha}}{\del x^M(\sigma)}\frac{\del \sigma^{\beta}}{\del x^N(\sigma)} \del_{\alpha} \tilde{\phi}(\sigma) \del_{\beta} \tilde{\phi}(\sigma) + V( \tilde{\phi}(\sigma)) \right), \label{scalar action tilde}
\eeq
where  $\sigma^{\alpha}=(\tau, \sigma^i)$   ($\alpha = 0,1,2,3,\; i=1,2,3,\;
\sigma^0=\tau$). 
In this action, let both the induced scalar $\tilde{\phi}(\sigma)$ and 
the coordinates $x^M(\sigma)$ be dynamical variables as in \cite{Dirac}.
We regard $\tau$ as time in this system.
Moreover, let the boundary values of the $x^M(\sigma)$ parametrize 
the same space-like hypersurface,
\beq
x^M(T,\sigma^i)=X^M(\sigma^i), 
\eeq
where $T$ is the boundary value of $\tau$ that is the final time. 
$\tilde{I}$ is invariant under the reparametrization of $\sigma$ by which both $\tilde{\phi}(\sigma)$ and the $x^M (\sigma)$ are transformed as scalars. If this diffeomorphism is fixed by the gauge fixing condition $x^M (\sigma)=\sigma^M$, $\tilde{I}$ reduces to $I$. Therefore $\tilde{I}$ and $I$ are equivalent. We will actually see below that the on-shell action of $\tilde{I}$ is equal to that of $I$.

The equations of motions of $\tilde{I}$ are
\beqa
&&\del_{\gamma} \left[ \mbox{det} \left( \frac{\del x}{\del \sigma} \right)  \left( \frac{\del \sigma^{\gamma}}{\del x^L} \left( \frac{1}{2}\eta^{MN} \frac{\del \sigma^{\alpha}}{\del x^M}\frac{\del \sigma^{\beta}}{\del x^N} \del_{\alpha} \tilde{\phi} \del_{\beta} \tilde{\phi} + V( \tilde{\phi}) \right) - \frac{\del \sigma^{\alpha}}{\del x^L} \eta^{MN} \frac{\del \sigma^{\gamma}}{\del x^M}\frac{\del \sigma^{\beta}}{\del x^N} \del_{\alpha} \tilde{\phi} \del_{\beta} \tilde{\phi}\right) \right]=0, \n
&&\del_{\gamma} \left( \mbox{det} \left( \frac{\del x}{\del \sigma} \right) \eta^{MN} \frac{\del \sigma^{\gamma}}{\del x^M}\frac{\del \sigma^{\beta}}{\del x^N} \del_{\beta} \tilde{\phi} \right) - \mbox{det} \left( \frac{\del x}{\del \sigma} \right) V'(\tilde{\phi})=0. \label{e.o.m. of scalar tilde }
\eeqa
Let us consider the classical solution $\bar{x}^M(\sigma)$ and 
$\bar{\tilde{\phi}}(\sigma)$ which satisfies the boundary condition,
\beq
\bar{x}^M(T,\sigma^i)=X^M(\sigma^i), \qquad
\bar{\tilde{\phi}}(T,\sigma^i)=\tilde{\Phi}(\sigma^i) \equiv \Phi(X(\sigma^i)). \label{boundary condition of scalar tilde} 
\eeq
By substituting this classical solution into $\tilde{I}$, we obtain the on-shell action,
\beq
\tilde{S}(T, \tilde{\Phi}, X)=-\int^T_{T_0} d \tau \int d^3 \sigma^i \mbox{det} \left( \frac{\del \bar{x}(\sigma)}{\del \sigma} \right) \left( \frac{1}{2}\eta^{MN} \frac{\del \sigma^{\alpha}}{\del \bar{x}^M(\sigma)}\frac{\del \sigma^{\beta}}{\del \bar{x}^N(\sigma)} \del_{\alpha} \bar{\tilde{\phi}}(\sigma) \del_{\beta} \bar{\tilde{\phi}}(\sigma) + V( \bar{\tilde{\phi}}(\sigma)) \right), \label{scalar on-shell tilde}
\eeq
where $T_0$ is the initial time. If we define $\bar{\phi}(x)$ by $\bar{\tilde{\phi}}(\sigma) = \bar{\phi}(\bar{x}(\sigma))$ and $x = \bar{x}(\sigma)$, $\bar{\phi}(x)$ satisfies (\ref{e.o.m. of scalar}) and (\ref{boundary condition}). Therefore the coordinate transformation 
$x = \bar{x}(\sigma)$ gives the desired relation
\beq
S(X, \Phi(X)) = \tilde{S}(T, \tilde{\Phi}, X).
\eeq

Let us derive the Hamilton-Jacobi equations satisfied by $ \tilde{S}$. By solving these equations we obtain $ \tilde{S}$ and hence $S$ as well. $\tilde{S}$ is a functional of the final time $T$ and the boundary values, $\tilde{\Phi} $ and $X$. The variation of $\tilde{S}$ with respect to $T$, $\tilde{\Phi}$, and $X$ is given by
\beqa
\delta \tilde{S}&=&-\int d^3 \sigma^i \left.\mbox{det} \left( \frac{\del \bar{x}}{\del \sigma} \right) \left( \frac{1}{2}\eta^{MN} \frac{\del \sigma^{\alpha}}{\del \bar{x}^M}\frac{\del \sigma^{\beta}}{\del \bar{x}^N} \del_{\alpha} \bar{\tilde{\phi}} \del_{\beta} \bar{\tilde{\phi}} + V( \bar{\tilde{\phi}}) \right) \delta T \right|_{\tau = T} \n
&&-\int d^3\sigma^i\mbox{det}  \left. \left( \frac{\del \bar{x}}{\del \sigma} \right) \eta^{MN} \frac{\del \tau}{\del \bar{x}^M}\frac{\del \sigma^{\beta}}{\del \bar{x}^N} \del_{\beta} \bar{\tilde{\phi}} \delta \bar{\tilde{\phi}} \right|_{\tau = T} \n
&&-\int d^3 \sigma^i\mbox{det} \left( \frac{\del \bar{x}}{\del \sigma} \right)  \left( \frac{\del \tau}{\del \bar{x}^L} \left( \frac{1}{2}\eta^{MN} \frac{\del \sigma^{\alpha}}{\del \bar{x}^M}\frac{\del \sigma^{\beta}}{\del \bar{x}^N} \del_{\alpha} \bar{\tilde{\phi}} \del_{\beta} \bar{\tilde{\phi}}  + V( \bar{\tilde{\phi}}) \right) \right. \n
&& \qquad\qquad\qquad\qquad\quad \left. \left. - \frac{\del \sigma^{\alpha}}{\del \bar{x}^L} \eta^{MN} \frac{\del \tau}{\del \bar{x}^M}\frac{\del \sigma^{\beta}}{\del \bar{x}^N} \del_{\alpha} \bar{\tilde{\phi}} \del_{\beta} \bar{\tilde{\phi}}\right) \delta \bar{x}^L \right|_{\tau = T}. \nonumber \label{delta S tilde}
\eeqa
We have used the equations of motion (\ref{e.o.m. of scalar tilde }) 
and $\delta \bar{\tilde{\phi}}(T_0, \sigma^i)=\delta \bar{x} (T_0, \sigma^i) = 0$. 
Moreover, the condition (\ref{boundary condition of scalar tilde}) implies
\beqa
\delta \bar{x}^M(T, \sigma^i)&=&\delta X^M(\sigma^i) - \del_{\tau} \bar{x}^M(T, \sigma^i) \delta T, \n
\delta \bar{\tilde{\phi}}(T, \sigma^i)&=&\delta \tilde{\Phi}(\sigma^i) - \del_{\tau} \bar{\tilde{\phi}}(T, \sigma^i) \delta T.
\eeqa
Then,
\beqa
\frac{\del \tilde{S}}{\del T}&=& 0, \n
\frac{\delta \tilde{S}}{\delta X^L}&=&  -\mbox{det} \left( \frac{\del \bar{x}}{\del \sigma} \right)  \left( \frac{\del \tau}{\del \bar{x}^L} \left( \frac{1}{2}\eta^{MN} \frac{\del \sigma^{\alpha}}{\del \bar{x}^M}\frac{\del \sigma^{\beta}}{\del \bar{x}^N} \del_{\alpha} \bar{\tilde{\phi}} \del_{\beta} \bar{\tilde{\phi}} + V( \bar{\tilde{\phi}}) \right) \right. \n
&& \qquad\qquad\qquad\qquad\qquad\; \left.\left. - \frac{\del \sigma^{\alpha}}{\del \bar{x}^L} \eta^{MN} \frac{\del \tau}{\del \bar{x}^M}\frac{\del \sigma^{\beta}}{\del \bar{x}^N} \del_{\alpha} \bar{\tilde{\phi}} \del_{\beta} \bar{\tilde{\phi}}\right)\right|_{\tau = T}, \n
\frac{\delta \tilde{S}}{\delta \tilde{\Phi}}&=& \left. -\mbox{det} \left( \frac{\del \bar{x}}{\del \sigma} \right) \eta^{MN} \frac{\del \tau}{\del \bar{x}^M}\frac{\del \sigma^{\beta}}{\del \bar{x}^N} \del_{\beta} \bar{\tilde{\phi}}\right|_{\tau = T}. \label{pre H-J of scalar tilde}
\label{derivatives of Stilde}
\eeqa
Here we define $L_M$ and $\gamma_{ij}$ for convenience:
\beqa
&&\left. L_M \equiv \epsilon_{M L_1 L_2 L_3} \del_1 X^{L_1} \del_2 X^{L_2} \del_3 X^{L_3}=\mbox{det}\left(\frac{\del \bar{x}}{\del \sigma}\right)
\frac{\del \tau}{\del \bar{x}^M}\right|_{\tau=T} \quad (\epsilon_{0123}=1), \n
&&\gamma_{ij} \equiv \del_i X^M  \del_j X^N  \eta_{MN}, \qquad \gamma \equiv
 \mbox{det}(\gamma_{ij}). 
\label{definitions}
\eeqa
$\mbox{}$From (\ref{derivatives of Stilde}) and (\ref{definitions}),
noting that 
\beqa
&&\del_i X^M L_M=0, \;\;\; \eta^{MN}L_M L_N=-\gamma, \n
&&\left. \frac{\del \sigma^{\alpha}}{\del \bar{x}^M} 
\del_{\alpha}\bar{\tilde{\phi}}\right|_{\tau=T}=
\frac{1}{\gamma}\frac{\delta \tilde{S}}{\delta \tilde{\Phi}}L_M 
+\gamma^{ij} \del_i \tilde{\Phi} \eta_{MN} \del_j X^N,
\eeqa
we obtain the Hamilton-Jacobi equations
\beqa
\frac{\delta \tilde{S}}{\delta X^M}&=&-\left( \frac{1}{2} \left( \frac{1}{\sqrt{\gamma }}\frac{\delta \tilde{S}}{\delta \tilde{\Phi}} \right)^2 + \frac{1}{2} \gamma^{ij}  \del_i \tilde{\Phi}\del_j \tilde{\Phi} + V(\tilde{\Phi}) \right)L_M - \gamma^{ij}  \del_i \tilde{\Phi} \eta_{MN} \del_j X^{N} \frac{\delta \tilde{S}}{\delta \tilde{\Phi}}\label{H-J 1}, \\
\frac{\del \tilde{S}}{\del T}&=&0 \label{H-J 2}.
\eeqa

It is instructive to rederive the above Hamilton-Jacobi equations in the canonical formalism following Dirac \cite{Dirac}. $\tilde{I}$ is rewritten in the canonical formalism as follows:
\beq
\tilde{I} = \int d \tau d^3 \sigma^i \left( P_M \del_{\tau} x^M + P_{\tilde{\phi}} \del_{\tau} \tilde{\phi} - C^M \Sigma_M( x, \tilde{\phi}, P_L, P_{\tilde{\phi}}) \right), \label{canonical}
\eeq
where
\beq
\Sigma_M( x, \tilde{\phi}, P_L, P_{\tilde{\phi}})= P_M + \left( \frac{1}{2} \left( \frac{1}{\sqrt{\gamma}}P_{\tilde{\phi}} \right)^2 + \frac{1}{2} \gamma^{ij} \del_i \tilde{\phi} \del_j \tilde{\phi} + V(\tilde{\phi}) \right) L_M + P_{\tilde{\phi}}\del_i \tilde{\phi}\gamma^{ij} \eta_{MN} \del_j x^N, \label{constraint}
\eeq
and $C^M$ are Lagrange multipliers. The constraints $\Sigma_M=0$ are first-class ones \footnote{Indeed, the Poisson brackets between the $\Sigma_M$
are $\{ \Sigma_M (\tau, \sigma^i), \Sigma_N(\tau, \sigma'^i) \} =0$. } and come from the invariance under the reparametrization of $\sigma$.
By applying the relation between the on-shell action and the canonical momenta,
\beq
P_M = \frac{\delta \tilde{S}}{\delta X^M}, \quad P_{\tilde{\phi}} = \frac{\delta \tilde{S}}{\delta \tilde{\phi}},
\eeq
to this constraint, we obtain (\ref{H-J 1}). (See section 2 in \cite{ST2}.)
 On the other hand, the ordinary Hamilton-Jacobi equation $\frac{\delta \tilde{S}}{\delta T} + H = 0$ reduces to (\ref{H-J 2}), since the Hamiltonian $H=\int d^3 \sigma^i C^M \Sigma_M$ vanishes on shell. 


\section{Application to gravitational systems}
\setcounter{equation}{0}
In this section, we consider the same problem as the previous section in gravitational systems. Let us consider as an example a four-dimensional gravity given by
\beq
I = \int d^4 x \sqrt{-g(x)} \left( R(x) -\frac{1}{2} g^{MN}(x) \del_M \phi(x) \del_N \phi(x) + V(\phi(x)) \right). \label{the action of gravity}
\eeq 
As before we consider the classical solution $\bar{g}_{MN}(x)$ and $\bar{\phi}(x)$ which satisfies the boundary condition
\beqa
&&\frac{ \del X^M (\sigma^k) }{ \del \sigma^i } \frac{ \del X^N (\sigma^k)}{\del \sigma^j }\bar{g}_{MN}(X(\sigma^k)) = \frac{ \del X^M (\sigma^k) }{ \del \sigma^i } \frac{ \del X^N (\sigma^k)}{\del \sigma^j } G_{MN}(X(\sigma^k)), \n
&&\bar{\phi}(X(\sigma^k))= \Phi(X(\sigma^k)), \label{the boundary condition of gravity} 
\label{bc}
\eeqa 
where the $X^M(\sigma)$ parametrize an space-like hypersurface.\footnote{In fact,
we must add the Gibbons-Hawking term to the action in order to impose
a boundary condition such as (\ref{bc}) consistently. However, the following
argument is still valid after adding this term.}
Note that this boundary condition is invariant under general coordinate transformations of $x$. We also consider the on-shell action
\beq
S(X, G(X), \Phi(X)) = \int d^4 x \sqrt{-\bar{g}(x)} \left( \bar{R}(x) -\frac{1}{2} \bar{g}^{MN}(x) \del_M \bar{\phi}(x) \del_N \bar{\phi}(x) + V( \bar{\phi}(x)) \right), \label{on-shell action of gravity}
\eeq
which is an analogue of (\ref{scalar on-shell action }).

Note that we can obtain (\ref{scalar action tilde}) from (\ref{I}) by performing a general coordinate transformation $x=x(\sigma)$.  
Similarly, we obtain from (\ref{the action of gravity})
\beq
\tilde{I} = \int d^4 \sigma \sqrt{-\tilde{g} (\sigma)} \left( \tilde{R}(\sigma)-\frac{1}{2} \tilde{g}^{\alpha\beta}(\sigma)\del_{\alpha} \tilde{\phi}(\sigma)\del_{\beta} \tilde{\phi}(\sigma) + V( \tilde{\phi}(\sigma)) \right). \label{action of gravity tilde}
\eeq
Note that $\tilde{I}$ does not depend on the $x^M(\sigma)$ and takes the same form as $I$, because $I$ is invariant 
under the general coordinate transformations. Now we consider the classical solution $\bar{\tilde{g}}_{\alpha\beta}(\sigma)$ and $\bar{\tilde{\phi}}(\sigma)$ which satisfies the boundary condition on the fixed-time hypersurface,
\beqa
&&\bar{\tilde{g}}_{ij}(T,\sigma^k) = \tilde{G}_{ij}(\sigma^k) \equiv \frac{\del \bar{x}^M(T, \sigma^k)}{\del \sigma^i} \frac{\del \bar{x}^N(T, \sigma^k)}{\del \sigma^j} G_{MN}(X(\sigma^k)), \n
&&\bar{\tilde{\phi}}(T,\sigma^k) = \tilde{\Phi}(\sigma^k) \equiv \Phi(X(\sigma^k)), \n
&&\bar{x}^M (T,\sigma^k) = X^M(\sigma^k), \label{boundary condition of gravity tilde}
\eeqa
Then the on-shell action of $\tilde{I}$ is
\beq
\tilde{S}(T, \tilde{G}, \tilde{\Phi}) = \int_{T_0}^T d \tau \int d^3 \sigma^i \sqrt{-\bar{\tilde{g}} (\sigma)} \left( \bar{\tilde{R}}(\sigma)-\frac{1}{2} \bar{\tilde{g}}^{\alpha\beta}(\sigma)\del_{\alpha} \bar{\tilde{\phi}}(\sigma)\del_{\beta} \bar{\tilde{\phi}}(\sigma) + V( \bar{\tilde{\phi}}(\sigma))\right).  
\eeq 
As in the previous section, if we define $\bar{g}_{MN}(x)$ and $\bar{\phi}(x)$ by $\bar{g}_{MN}(\bar{x}(\sigma))= \frac{\del \sigma^{\alpha}}{\del \bar{x}^M}\frac{\del \sigma^{\beta}}{\del \bar{x}^N}\bar{\tilde{g}}_{\alpha\beta}(\sigma)$, $\bar{\phi}(\bar{x}(\sigma))=\bar{\tilde{\phi}}(\sigma)$ and $x=\bar{x}(\sigma)$, $\bar{g}_{MN}(x)$ and $\bar{\phi}(x)$ satisfy the equation of motion of $I$ and the boundary condition (\ref{the boundary condition of gravity}). It follows again that the on-shell actions are equivalent:
\beq
S(X, G(X), \Phi(X)) = \tilde{S}(T, \tilde{G}, \tilde{\Phi}).
\eeq
Contrary to the case of the flat space, the Hamilton-Jacobi equations of $\tilde{I}$ clearly take the same forms as the ordinary ones of $I$, which are satisfied by the on-shell action with the boundary values of the fields given on the fixed-time hypersurface. Therefore if one knows the on-shell action with the boundary values of the fields given on the fixed-time hypersurface, one can obtain the on-shell action with those given on an arbitrary space-like hypersurface. 
Note that this consequence directly comes from the facts that $I$ is invariant
under general coordinate transformations and that an arbitrary space-like
hypersurface can be transformed to a fixed-time hypersurface by a general
coordinate transformation.

Finally we apply the above argument to our results in \cite{ST, ST2}. We reported in \cite{ST, ST2} that the D-brane and M-brane effective actions are on-shell actions of supergravities, which are defined on fixed-radial-time hypersurfaces. Obviously, the above argument holds even when we replace the time with the radial time that is a space-like direction. It follows that the D-brane and M-brane effective actions given by arbitrary embedding functions are also on-shell actions in supergravities. We take the D3-brane case as an example below and write down the explicit form of the on-shell action. The on-shell actions corresponding to the other branes can be written down explicitly in the same way. 

We start with the five-dimensional gravity which we obtained in \cite{ST, ST2} by reducing type IIB supergravity on $S^5$,
\beqa
&&I_5 =  \int d^5 x \sqrt{-h} 
\left[e^{-2\phi+\frac{5}{4}\rho}
\left( R +4\partial_{\alpha}\phi\partial^{\alpha}\phi 
+\frac{5}{4}\partial_{\alpha}\rho\partial^{\alpha}\rho
-5\partial_{\alpha}\phi\partial^{\alpha}\rho 
-\frac{1}{2}|H_3|^2 \right) \right.\n
&& \qquad \qquad \qquad \quad \left.
-\frac{1}{2}e^{\frac{5}{4}\rho} \left(
 |F_1|^2
+|F_3+c_0 H_3|^2+|F_{5}+c_2 H_3|^2 \right) 
+e^{-2\phi+\frac{3}{4}\rho}R^{(S^5)} \right],
\label{5Daction}
\eeqa
where $H_3=d b_2$, $F_{p+2}=d c_{p+1}$ $(p=-1,1,3)$. The $x^M$ $(M=0,\cdots,4)$ are five-dimensional coordinates. $h_{MN}$, $\phi$, $\rho$, $b_2$ and $c_{p+1}$ are the five-dimensional metric, the dilaton, the warp factor of $S^5$, the Kalb-Ramond field and the R-R $(p+1)$-form, respectively. $R^{(S^5)}$ is the constant curvature of $S^5$. 

In this case, we are interested in on-shell actions with the boundary values
of fields given on general {\em time-like} hypersurfaces, since we
interpret these hypersurfaces as the worldvolumes of D3-brane.
Let $X^M(\sigma^{\mu})\;\;(\mu=0,\cdots,3)$ be embedding functions 
satisfying such a hypersurface. We consider the classical solution
$\bar{h}_{MN}(x)$, $\bar{\phi}(x)$, $\bar{\rho}(x)$, $\bar{b}_{MN}(x)$ and
$\bar{c}_{M_1 \cdots M_{p+1}}(x)$ which
satisfies the boundary condition
\beqa
&&\frac{\pa X^M(\sigma^{\lambda})}{\pa \sigma^{\mu}}
\frac{\pa X^N(\sigma^{\lambda})}{\pa \sigma^{\nu}}
\bar{h}_{MN}(X(\sigma^{\lambda}))
=\frac{\pa X^M(\sigma^{\lambda})}{\pa \sigma^{\mu}}
\frac{\pa X^N(\sigma^{\lambda})}{\pa \sigma^{\nu}}
G_{MN}(X(\sigma^{\lambda})), \n
&&\bar{\phi}(X(\sigma^{\mu}))=\Phi(X(\sigma^{\mu})), \n
&&\bar{\rho}(X(\sigma^{\mu}))=\Upsilon(X(\sigma^{\mu})), \n
&&\frac{\pa X^M(\sigma^{\lambda})}{\pa \sigma^{\mu}}
\frac{\pa X^N(\sigma^{\lambda})}{\pa \sigma^{\nu}}
\bar{b}_{MN}(X(\sigma^{\lambda}))
=\frac{\pa X^M(\sigma^{\lambda})}{\pa \sigma^{\mu}}
\frac{\pa X^N(\sigma^{\lambda})}{\pa \sigma^{\nu}}
B_{MN}(X(\sigma^{\lambda})), \n
&&\frac{\pa X^{M_{1}}(\sigma^{\nu})}{\pa \sigma^{\mu_1}} \cdots
\frac{\pa X^{M_{p+1}}(\sigma^{\nu})}{\pa \sigma^{\mu_{p+1}}} 
\bar{c}_{M_1\cdots M_{p+1}} (X(\sigma^{\nu})) \n
&&=\frac{\pa X^{M_{1}}(\sigma^{\nu})}{\pa \sigma^{\mu_1}} \cdots
\frac{\pa X^{M_{p+1}}(\sigma^{\nu})}{\pa \sigma^{\mu_{p+1}}} 
C_{M_1\cdots M_{p+1}}(X(\sigma^{\nu})).
\eeqa
As before $\tilde{I}_5$ is obtained by replacing the fields
in $I_5$ with those with tilde, 
$\tilde{h}_{\alpha\beta}(\sigma^{\alpha})$, $\tilde{\phi}(\sigma^{\alpha})$, $\tilde{\rho}(\sigma^{\alpha})$, $\tilde{b}_2(\sigma^{\alpha})$ and $\tilde{c}_{p+1}(\sigma^{\alpha})$, where $\sigma^{\alpha}=(\sigma^{\mu}, \sigma^4)$ $(\mu=0,\cdots,3)$. 
We regard $\sigma^4$ as time, and 
denote the boundary value of $\sigma^4$ by U. 
The classical solution of $\tilde{I}_5$, 
$\bar{x}^M(\sigma^{\alpha})$,
$\bar{\tilde{h}}_{\alpha\beta}(\sigma^{\gamma})$, 
$\bar{\tilde{\phi}}(\sigma^{\alpha})$, $\bar{\tilde{\rho}}(\sigma^{\alpha})$, 
$\bar{\tilde{b}}_{\alpha\beta}(\sigma^{\gamma})$ and 
$\bar{\tilde{c}}_{\alpha_1 \cdots \alpha_{p+1}}(\sigma^{\beta})$,
corresponding to the above solution of $I_5$ satisfies the boundary condition  
\beqa
&&\bar{x}^M (\sigma^{\mu}, U) = X^M(\sigma^{\mu}),  \n
&&\bar{\tilde{h}}_{\mu\nu}(\sigma^{\lambda},U)
= \tilde{G}_{\mu\nu}(\sigma^{\lambda})
\equiv \frac{ \del X^M (\sigma^{\lambda})}{ \del \sigma^{\mu} } \frac{ \del X^N (\sigma^{\lambda})}{\del \sigma^{\nu} } G_{MN}(X(\sigma^{\lambda})),\n
&&\bar{\tilde{\phi}}(\sigma^{\mu},U)=\tilde{\Phi}(\sigma^{\mu})
\equiv \Phi(X(\sigma^{\mu})) ,\n
&&\bar{\tilde{\rho}}(\sigma^{\mu},U)=\tilde{\Upsilon}(\sigma^{\mu})
\equiv \Upsilon(X(\sigma^{\mu})) ,\n
&&\bar{\tilde{b}}_{\mu\nu}(\sigma^{\lambda},U)
= \tilde{B}_{\mu\nu}(\sigma^{\lambda})
\equiv \frac{ \del X^M (\sigma^{\lambda})}{ \del \sigma^{\mu} } \frac{ \del X^N (\sigma^{\lambda})}{\del \sigma^{\nu} } B_{MN}(X(\sigma^{\lambda})),\n
&&\bar{\tilde{c}}_{\mu_1 \cdots \mu_{p+1}}(\sigma^{\nu},U)=
\tilde{C}_{\mu_1 \cdots \mu_{p+1}}(\sigma^{\nu}) 
\equiv \frac{\pa X^{M_{1}}(\sigma^{\nu})}{\pa \sigma^{\mu_1}} \cdots
\frac{\pa X^{M_{p+1}}(\sigma^{\nu})}{\pa \sigma^{\mu_{p+1}}} 
C_{M_1\cdots M_{p+1}}(X(\sigma^{\nu})). \n
\label{the boundary condition of supergravity} 
\eeqa

In order to derive the H-J equations, we perform 
the ADM decomposition as follows:
\beqa
ds^2_5&=&h_{\alpha\beta}d\sigma^{\alpha} d\sigma^{\beta} \n
&=&(n^2 + n^{\mu}n_{\mu})(d\sigma^4 )^2
+2 n_{\mu} d\sigma^{\mu} d\sigma^4+h_{\mu\nu}d\sigma^{\mu} d\sigma^{\nu},
\eeqa
where $n$ and $n_{\mu}$ are the lapse function and the shift functions, 
respectively.
By adding the Gibbons-Hawking term, $\tilde{I}_5$ can be rewritten 
in the canonical form as 
\beqa
&&\tilde{I}_5 = \int d^5 \sigma 
\sqrt{-h}(\pi^{\mu\nu}\pa_{\sigma^4} \tilde{h}_{\mu\nu}
+\pi_{\tilde{\phi}}\pa_{\sigma^4} \tilde{\phi}
+\pi_{\tilde{\rho}}\pa_{\sigma^4} \tilde{\rho}
+\pi_{\tilde{b}_2}^{\mu\nu} \pa_{\sigma^4} \tilde{b}_{\mu\nu} 
+\sum_p \pi_{\tilde{c}_{p+1}}^{\mu_1 \cdots \mu_{p+1}} \pa_{\sigma^4}
\tilde{c}_{\mu_1 \cdots \mu_{p+1}} \n
&&\qquad \qquad \qquad    \;\;\;\;\;
-nH-n_{\mu}H^{\mu}-\tilde{b}_{4\mu}Z_{\tilde{b}_2}^{\mu}
-\tilde{c}_{4\mu}Z_{\tilde{c}_2}^{\mu}
-\tilde{c}_{4\mu\nu\lambda}Z_{\tilde{c}_4}^{\mu\nu\lambda}),
\label{ADM}
\eeqa
with
\beqa
&&H=-e^{2\tilde{\phi}-\frac{5}{4}\tilde{\rho}} \left( (\pi^{\mu\nu})^2
+\frac{1}{2}{\pi_{\tilde{\phi}}}^2
+\frac{1}{2}\pi^{\mu}_{\;\; \mu}\pi_{\tilde{\phi}}
+\frac{4}{5}{\pi_{\tilde{\rho}}}^2+\pi_{\tilde{\phi}}\pi_{\tilde{\rho}} 
+\left( \pi_{\tilde{b}_2}^{\mu\nu}-\tilde{c}\pi_{\tilde{c}_2}^{\mu\nu}
-6\tilde{c}_{\lambda\rho}\pi_{\tilde{c}_4}^{\mu\nu\lambda\rho} 
\right)^2 \right) \n
&& \qquad -e^{-\frac{5}{4}\tilde{\rho}} \left( \frac{1}{2}{\pi_{\tilde{c}_0}}^2
+(\pi_{\tilde{c}_2}^{\mu\nu})^2 +12(\pi_{\tilde{c}_4}^{\mu\nu\lambda\rho})^2  
\right)
- {\cal L}, \\
&&H^{\mu}=-2\nabla_{\nu}\pi^{\mu\nu}+\pi_{\tilde{\phi}}\pa^{\mu}\tilde{\phi}
+\pi_{\tilde{\rho}}\pa^{\mu}\tilde{\rho}
 +\pi_{\tilde{b}_2 \nu\lambda} \tilde{H}^{\mu\nu\lambda} \n
&& \qquad \;\; +\pi_{\tilde{c}_0}\pa^{\mu}\tilde{c}
+\pi_{\tilde{c}_2 \nu\lambda}\tilde{F}^{\mu\nu\lambda}
+\pi_{\tilde{c}_4 \nu\lambda\rho\sigma}(\tilde{F}^{\mu\nu\lambda\rho\sigma}
+4\tilde{c}^{\mu\nu}\tilde{H}^{\lambda\rho\sigma}),\\
&&Z_{\tilde{b}_2}^{\mu}=2\nabla_{\nu}\pi_{\tilde{b}_2}^{\mu\nu}, \\
&&Z_{\tilde{c}_2}^{\mu}=2\nabla_{\nu}\pi_{\tilde{c}_2}^{\mu\nu}
-4\pi_{\tilde{c}_4}^{\mu\nu\lambda\rho}\tilde{H}_{\nu\lambda\rho}, \\
&&Z_{\tilde{c}_4}^{\mu\nu\lambda}
=4\nabla_{\rho}\pi_{\tilde{c}_4}^{\mu\nu\lambda\rho},
\eeqa
where 
\beqa
&&h=\det h_{\mu\nu}, \n
&&\tilde{H}_{\mu\nu\lambda}=3\pa_{[\mu}\tilde{b}_{\nu\lambda]}, \;\;\;
\tilde{F}_{\mu_1 \cdots \mu_{p+1}}
=(p+1)\pa_{[\mu_1}\tilde{c}_{\mu_2 \cdots \mu_{p+2}]}, \n
&&{\cal L}=e^{-2\tilde{\phi}+\frac{5}{4}\tilde{\tilde{\rho}}} 
\left(R^{(4)}+4\nabla_{\mu}\nabla^{\mu}\tilde{\phi}
-\frac{5}{2}\nabla_{\mu}\nabla^{\mu}\tilde{\rho}
-4\pa_{\mu}\tilde{\phi}\pa^{\mu}\tilde{\phi}
-\frac{15}{8}\pa_{\mu}\tilde{\rho}\pa^{\mu}\tilde{\rho}
+5\pa_{\mu}\tilde{\phi}\pa^{\mu}\tilde{\rho}
-\frac{1}{12}\tilde{H}_{\mu\nu\lambda}\tilde{H}^{\mu\nu\lambda} \right) \n
&& \qquad +e^{\frac{5}{4}\tilde{\rho}} 
\left( -\frac{1}{2}\pa_{\mu}\tilde{c}\pa^{\mu}\tilde{c}
-\frac{1}{12}(\tilde{F}_{\mu\nu\lambda}+\tilde{c}\tilde{H}_{\mu\nu\lambda})
(\tilde{F}^{\mu\nu\lambda}+\tilde{c}\tilde{H}^{\mu\nu\lambda}) \right) 
+e^{-2\tilde{\phi}+\frac{3}{4}\tilde{\rho}}R^{(S^5)}.
\eeqa
The H-J equations of this system are
\beqa
\frac{\pa \tilde{S}}{\pa U}=0, 
\label{H-J1}
\eeqa
and
\beqa
H=0,\;\;\; H^{\mu}=0, \;\;\; Z_{\tilde{b}_2}=0,\;\;\;
Z_{\tilde{c}_2}=0,\;\;\; Z_{\tilde{c}_4}=0
\label{H-J2}
\eeqa
with the following replacements:
\beqa
&&\tilde{h}_{\mu\nu} \rightarrow \tilde{G}_{\mu\nu},\;\;\;
\tilde{\phi} \rightarrow \tilde{\Phi},\;\;\;
\tilde{\rho} \rightarrow \tilde{\Upsilon},\;\;\;
\tilde{b}_{\mu\nu} \rightarrow \tilde{B}_{\mu\nu}, \;\;\;
\tilde{c}_{\mu_1 \cdots \mu_{p+1}} 
\rightarrow \tilde{C}_{\mu_1 \cdots \mu_{p+1}}, \n
&&\pi^{\mu\nu} \rightarrow \frac{1}{\sqrt{-\tilde{G}}}
\frac{\delta \tilde{S}}{\delta \tilde{G}_{\mu\nu}},\;\;\;
\pi_{\tilde{\phi}} \rightarrow \frac{1}{\sqrt{-\tilde{G}}}
\frac{\delta \tilde{S}}{\delta \tilde{\Phi}},\;\;\;
\pi_{\tilde{\rho}} \rightarrow \frac{1}{\sqrt{-\tilde{G}}}
\frac{\delta \tilde{S}}{\delta \tilde{\Upsilon}},\n
&&\pi_{\tilde{b}_2}^{\mu\nu} \rightarrow \frac{1}{\sqrt{-\tilde{G}}}
\frac{\delta \tilde{S}}{\delta \tilde{B}_{\mu\nu}},\;\;\;
\pi_{\tilde{c}_{p+1}}^{\mu_1 \cdots \mu_{p+1}} \rightarrow
\frac{1}{\sqrt{-\tilde{G}}}
\frac{\delta \tilde{S}}{\delta \tilde{c}_{\mu_1 \cdots \mu_{p+1}}}.
\label{pianddelS}
\eeqa
The H-J equation (\ref{H-J1}) implies that $\tilde{S}$ does not depend on the
final time explicitly while the last four equations in (\ref{H-J2})
imply that $\tilde{S}$ is invariant under the diffeomorphism in four dimensions
and the $U(1)$ gauge transformations for $\tilde{B}_{\mu\nu}$ and 
$\tilde{C}_{\mu_1 \cdots \mu_{p+1}}$. The first equation $H=0$ in (\ref{H-J2})
is only a nontrivial one that can be determine the form of $\tilde{S}$.

We solve the H-J equations under the condition that the fields in $\tilde{I}_5$
depend only on $\sigma^4$, and denote a solution to the H-J equations under
this condition by $\tilde{S}_0$. Then the H-J equation $H=0$ reduces to
\beqa
&&-e^{2\tilde{\Phi}-\frac{5}{4}\tilde{\Upsilon}} \left(
\left(\frac{1}{\sqrt{-\tilde{G}}}
\frac{\delta \tilde{S}_0}{\delta \tilde{G}_{\mu\nu}}\right)^2
+\frac{1}{2}\tilde{G}_{\mu\nu}
\frac{1}{\sqrt{-\tilde{G}}}\frac{\delta \tilde{S}_0}{\delta \tilde{G}_{\mu\nu}}
\frac{1}{\sqrt{-\tilde{G}}}\frac{\delta \tilde{S}_0}{\delta \tilde{\Phi}}
+\frac{1}{2}\left(\frac{1}{\sqrt{-\tilde{G}}}
\frac{\delta \tilde{S}_0}{\delta \tilde{\Phi}}\right)^2 \right. \n
&&\qquad\qquad+\frac{4}{5}\left(\frac{1}{\sqrt{-\tilde{G}}}
\frac{\delta \tilde{S}_0}{\delta \tilde{\Upsilon}}\right)^2 
+\frac{1}{\sqrt{-\tilde{G}}}
\frac{\delta \tilde{S}_0}{\delta \tilde{\Phi}}
\frac{1}{\sqrt{-\tilde{G}}}
\frac{\delta \tilde{S}_0}{\delta \tilde{\Upsilon}} \n
&&\qquad\qquad \left.
+\left(\frac{1}{\sqrt{-\tilde{G}}}
\frac{\delta \tilde{S}_0}{\delta \tilde{B}_{\mu\nu}}
-\tilde{C}\frac{1}{\sqrt{-\tilde{G}}}
\frac{\delta \tilde{S}_0}{\delta \tilde{C}_{\mu\nu}}
-6\tilde{C}_{\lambda\rho}\frac{1}{\sqrt{-\tilde{G}}}
\frac{\delta \tilde{S}_0}{\delta \tilde{C}_{\mu\nu\lambda\rho}}\right)^2 
\right) \n
&&-e^{-\frac{5}{4}\tilde{\Upsilon}} \left(
\frac{1}{2}\left(\frac{1}{\sqrt{-\tilde{G}}}
\frac{\delta \tilde{S}_0}{\delta \tilde{C}}\right)^2
+\left(\frac{1}{\sqrt{-\tilde{G}}}
\frac{\delta \tilde{S}_0}{\delta \tilde{C}_{\mu\nu}}\right)^2
+12\left(\frac{1}{\sqrt{-\tilde{G}}}
\frac{\delta \tilde{S}_0}{\delta \tilde{C}_{\mu\nu\lambda\rho}}
\right)^2 \right) \n
&&=e^{-2\tilde{\Phi}+\frac{3}{4}\tilde{\Upsilon}}R^{(S^5)}.
\label{HJ}
\eeqa

We showed in \cite{ST, ST2} that the following form is one of the solutions
to the reduced H-J equations.
\beq
\tilde{S}_0=\tilde{S}_c+\tilde{S}_{BI}+\tilde{S}_{WZ}
\label{S0}
\eeq
with
\beqa
\tilde{S}_c&=&\pm \sqrt{5R^{(S^5)}} \int d^4\sigma \sqrt{-\tilde{G}} e^{-2\tilde{\Phi}+\tilde{\Upsilon}}, \n
\tilde{S}_{BI}&=&\beta \int d^4\sigma e^{-\tilde{\phi}} \sqrt{-\det (\tilde{G}_{\mu\nu}+\tilde{\mathcal{F}}_{\mu\nu})}, \n
\tilde{S}_{WZ}&=&\pm\beta\left( \int \tilde{C_4} + \int \tilde{C_2} \wedge \tilde{\mathcal{F}} 
+ \frac{1}{2}\int \tilde{C_0} \tilde{\mathcal{F}} \wedge \tilde{\mathcal{F}} \right), 
\label{supergravity on-shell action}
\eeqa 
where $\tilde{\mathcal{F}}_{\mu\nu} = \tilde{B}_{\mu\nu} + F_{\mu\nu}$, both
$F_{\mu\nu}$ and $\beta$ are arbitrary constants and the signs in $\tilde{S}_c$
and $\tilde{S}_{WZ}$ can take all combinations. Thus $\tilde{S}_0$ is
an on-shell action of $\tilde{I}_5$. The argument in the first half 
of this section shows that one obtains the corresponding on-shell action $S_0$
of $I_5$
by rereading the fields with tilde in (\ref{S0}) using 
(\ref{the boundary condition of supergravity}). Hence we conclude that
the effective actions of D$p$-brane and M$p$-brane whose worldvolumes are
defined by $p+1$ embedding functions are on-shell actions of supergravities
reduced in $p+2$ dimensions. This is an important generalization of our
results in \cite{ST,ST2} although its derivation is rather simple.


\section{Discussion}
\setcounter{equation}{0}
As we have seen in the example of the D3-brane case in the previous section,
all we can do at present is to reduce ten-dimensional supergravities or
eleven-dimensional supergravity to $(p+2)$-dimensional gravity and obtain
the $p$-brane effective action, whose $(p+1)$-dimensional worldvolume 
is given by arbitrary embedding functions, as an on-shell action of the
$(p+2)$-dimensional gravity. In other words, we can only consider
hypersurfaces whose codimension is one.
Thus our results in this note are not completely
satisfactory, since in string theories one can consider D-branes 
whose codimension is larger than one. Hence, an issue we should next study 
is an `on-shell action' that one obtains when one takes a hypersurface 
whose codimension is larger than one, specifies
the `boundary' values on the hypersurface and 
substitutes into the action the classical solution satisfying the `boundary'
condition. We need to develop a formalism that gives such an `on-shell action'
and see whether the $p$-brane effective action is an `on-shell action' of
a $(p+k)$-dimensional gravity which is obtained by reducing ten-dimensional
supergravities or eleven-dimensional supergravity, 
where $2 < k \leq 10-p \;\;(\mbox{or}\;11-p)$.\footnote{In \cite{NO}, 
the Nambu-Goto action and its corrections were derived from four-dimensional 
field theories. Perhaps these works will 
give a hint to the above
issue.}

\vspace{0.5cm}

We would like to thank T. Yoneya for bringing 
our attention to Ref.\cite{Dirac}.
The work of M.S. is supported in part by 
Research Fellowships of the Japan Society for the Promotion of Science (JSPS) 
for Young Scientists (No.13-01193).


\end{document}